\documentclass[prd,unsortedaddress,superscriptaddress,a4paper,nofootinbib,%
nobalancelastpage,preprintnumbers,showpacs]{revtex4}
\usepackage{url}
\usepackage{dcolumn}
\usepackage{longtable}
\usepackage{latexsym,amsbsy}
\usepackage[dvips]{graphicx}
\renewcommand{\vec}[1]{\boldsymbol #1}
\newcommand{\sutf}{\ensuremath{\mathrm{SU(3)_{\text{F}}}}}
\newcommand{\suds}{\ensuremath{\mathrm{SU(2)_{\text{s}}}}}
\newcommand{\sutc}{\ensuremath{\mathrm{SU(3)_{\text{c}}}}}
\newcommand{\suscs}{\ensuremath{\mathrm{SU(6)_{\text{cs}}}}}

\newcommand{\jp}{\ensuremath{\mathrm{J}/\psi}}
\newcommand{\be}{\begin{equation}}
\newcommand{\ee}{\end{equation}}
\newcommand{\SpSp}[2]{ \ensuremath{\vec{\sigma }_{#1}.\vec{\sigma
}_{#2}}}
\newcommand{\lala}[2]{ \ensuremath{{\tilde{\lambda}_{#1}.
\tilde{\lambda}_{#2}}}}
\newcommand{\df}[2]{\ensuremath{ {\raise1pt\hbox{$\displaystyle
#1$}\over
\raise -2pt \hbox{$\displaystyle #2$}}}}
\begin{document}
\title{\bf Chromomagnetism, flavour symmetry\\ breaking
and S-wave tetraquarks}
\date{\today}
\author{F.~Buccella}
\email{buccella@na.infn.it}
\affiliation{Universit\`a  di Napoli Federico II,
Dipartimento di Scienze Fisiche
and INFN, Sezione di Napoli,\\
Complesso Universitario di Monte Sant' Angelo,
via Cintia
I-80126 Napoli, Italy}
\author{H.~H\o gaasen}
\email{hallstein.hogasen@fys.uio.no}
\affiliation{Department of Physics, University of Oslo,
Box 1048  NO-0316 Oslo, Norway}
\author{J.-M.~Richard}
\email{jean-marc.richard@lpsc.in2p3.fr}
\affiliation{Laboratoire de Physique Subatomique et Cosmologie,
 Universit\'e Joseph Fourier--IN2P3-CNRS\\[-1pt]
53, avenue des Martyrs, 38026  Grenoble cedex, France}
\author{P.~Sorba}
\email{sorba@lapp.in2p3.fr}
\affiliation{Laboratoire d'Annecy-le-Vieux de Physique Th\'eorique
(LAPTH)\\[-1pt]
9, chemin de Bellevue, B.P. 110, 74941 Annecy-le-Vieux Cedex, France}
\preprint{\begin{minipage}{2.7cm}
DSF/13(2006)\\
LAPTH-1150/06\\
LPSC-06-37\\
hep-ph/0608001
\end{minipage}
}
\begin{abstract}
The chromomagnetic interaction, with full account for flavour-symmetry
breaking, is applied to S-wave configurations containing two quarks
and two antiquarks.  Phenomenological implications are discussed for
light, charmed, charmed and strange, hidden-charm and double-charm
mesons, and extended to their analogues with beauty.
\end{abstract}
\maketitle
\section{Introduction}\label{se:int}
The question of the existence of multiquark hadrons beyond ordinary
mesons and baryons has been addressed since the beginning of the quark
model.  It has been particularly discussed recently with the firm or
tentative discovery of new hadron states in a variety of experiments.
For a review of recent results, see, e.g.,
Refs.~\cite{Klempt:2006uy,Barnes:2005zy,Jin:2004ck,Swanson:2006st}.

Different mechanisms have been proposed to form stable or metastable
multiquarks in the ground state.  The most natural mechanism,
especially for states close to a hadron--hadron threshold, is provided
by nuclear forces, extrapolated from the nucleon--nucleon interaction,
and acting between any pair of hadrons containing light quarks.  This
led several authors to predict the existence of
$\mathrm{D}\mathrm{D}^*$ and
$\mathrm{D}^*\overline{\mathrm{D}}({}+\mathrm{c.c.})$ molecules
\cite{Voloshin:1976ap,Tornqvist:1991ks,
Ericson:1993wy,Braaten:2005ai,Swanson:2003tb}.  According to these
authors (see, also,
\cite{Close:2003sg,Suzuki:2005ha,Kalashnikova:2005ui}), the latter
configuration is perhaps seen in the X(3872) \cite{Choi:2003ue},
though other interpretations have been proposed for this narrow meson
resonance with hidden charm \cite{Maiani:2004vq,Hogaasen:2005jv}.
Stable or metastable multicharmed dibaryons are also predicted in this
nuclear-physics type approach \cite{Julia-Diaz:2004rf}.

Flavour independence is a key property of QCD, at least in the
heavy-quark limit.  Quarks are coupled to the gluon field through
their colour, not their mass, and this induces a static interquark
potential which is independent of the flavour content, in the same way
as the same Coulomb interaction is kept acting on antiprotons, kaons,
muons and electrons when exotic atoms and molecules are studied
\cite{Armour:2005}.  The mechanism by which the hydrogen molecule is
more deeply bound than the positronium molecule remains valid,
\textsl{mutatis mutandis}, in hadron physics with flavour independence
and favours the binding of $(QQ\bar{q}\bar{q})$ below the threshold of
two heavy-flavoured mesons, when the quark-mass ratio $Q/q$ increases
\cite{Ader:1981db,Zouzou:1986qh,Heller:1986bt,Carlson:1988hh,Lipkin:1986dw,%
Brink:1998as,Gelman:2002wf,Janc:2004qn}.

The best known mechanism for multiquark binding is based on
spin-dependent forces.  In the late 70's, Jaffe
\cite{Jaffe:1976ig,Jaffe:1976ih} proposed a $(q^2\bar{q}^2)$ picture
of some scalar mesons, as a solution to the puzzle of their low mass,
decay and production properties, and abundance.  He also discovered
that the colour--spin operator entering the widely-accepted models
sometimes provides multiquark states with a coherent attraction which
is larger than the sum of the attractive terms in the decay products,
hence favouring the formation of bound states.  An example is the
so-called $\mathrm{H}$ dibaryon \cite{Jaffe:1976yi}, with spin $S=0$
and quark content $(ssuudd)$, tentatively below any threshold made of
two light baryons.  This prediction stimulated an intense experimental
work, which did not lead to any positive evidence, see, e.g.,
\cite{Yamamoto:2000wf}.

This model for the $\mathrm{H}$ also provoked much theoretical
activity.  New configurations were found, in which the chromomagnetic
effects are favourable, such as the 1987-vintage pentaquark
\cite{Gignoux:1987cn,Lipkin:1987sk}, $\mathrm{P}=(Q\bar{q}^4)$.  A
comprehensive systematics of multiquark configurations with favourable
chromomagnetic effects can be found, e.g., in
\cite{Silvestre-Brac:1992yg,Leandri:1989su,Lichtenberg:1992ji} and the
references therein.

 In the early days of multiquark investigations, chromomagnetic
 effects were also intensively used in an attempt to explain the
 narrow hadronic resonances which were observed at that time
 \cite{Montanet:1980te}.  Models were proposed for these hadrons, with
 two clusters of complementary colour separated by an orbital
 momentum, to prevent the decay into colour singlets without pair
 creation and thereby give possible long lived states
 \cite{Johnson:1975sg,Chan:1977st,Chan:1978nk,Hogaasen:1978jw}.  The
 effective mass of each cluster was computed from the chromomagnetic
 interaction and effective quark masses.  The most studied states were
 the tetraquarks called ``baryonium" and the pentaquarks called
 ``mesobaryonium".

In the limit of exact ${\mathrm{SU(N)_{\text{F}}}}$ flavour-symmetry,
the chromomagnetic model reads
\be\label{int:eq:cm1}
H=\sum_i m_i-
C\sum_{i<j}\tilde{\lambda}_i.\tilde{\lambda}_j\,\vec{\sigma}_i.\vec{\sigma}_j~,
\ee
where $\vec{\sigma}_i$ is the spin operator and $\tilde{\lambda}_{i}$
the colour operator acting on the $i^{\text{th}}$ quark, and each
effective mass $m_i$ includes the constituent quark mass and its
chromoelectric energy (binding effect).  There is already an abundant
literature on how to estimate the expectation value of the
chromomagnetic operator for multiquark configurations, in particular
using some powerful group-theoretical techniques.  The Hamiltonian
(\ref{int:eq:cm1}) is expressed in terms of Casimir operators of the
spin, colour and spin--colour groups.  When the overall strength
factor $C$ is replaced by a coupling $C_{ij}$ which depends on the
quark flavour, an explicit basis is required to estimate the
eigenstates of $H$.

Note that the role of \sutf\ symmetry breaking has already been
analysed in the literature, see, in particular,
\cite{Rosner:1985yh,Karl:1987cg,Karl:1987uf,Sakai:1999qm} for the H
and the P. It often happens that the corrections weaken or even spoil
the binding predicted in the \sutf\ limit.

In this paper, a detailed formalism is presented to fully account for
 flavour-symmetry breaking in the chromomagnetic interaction, and an
 application is given to the sector of systems made of two quarks and
 two antiquarks in a relative S-wave, i.e., scalar ($J^P=0^+$), axial
 ($1^+$) and tensor ($2^+$) mesons. The question then is how to
 extrapolate the strength of the chromomagnetic interaction from the
 meson or baryon sector to the case of multiquark configurations.

There has been several investigations of multiquark states
using the remarkable know-how of few-body physics. The strategy here
consists in writing down an explicit Hamiltonian with kinetic energy
operator, spin-independent confining forces and spin-dependent
terms, tuning the parameters to reproduce some known mesons and
baryons, and solve the multiquark problem. This involves an
extrapolation of the linear quark--antiquark potential toward
multiquark states and an \textsl{ad-hoc} regularisation of the
contact interaction,  which then can be treated beyond first order.

The present approach is somewhat complementary.  The role of the
chromomagnetic interaction is analysed from the point of view of the
symmetry properties, to deduce patterns shared by a whole class of
models.  The study is restricted to the chromomagnetic model, though
it has been challenged recently by models where the hyperfine
splittings of hadrons is described by instanton-induced forces or
spin-flavour terms.  The multiquark sector in these models is reviewed
by Stancu \cite{Stancu:1999xc} or Sakai et al.\ \cite{Sakai:1999qm}.

It is well known that a colour singlet configuration with two quarks
and two antiquarks has at least one component which is a product of
two colour singlets.  Hence most states are very broad, since unstable
against spontaneous dissociation, and give only indirect signatures.
However, in rare circumstances, the dissociation is kinematically
suppressed, resulting into a remarkably small width.  This is the
scenario proposed recently for the X(3872) \cite{Hogaasen:2005jv}.

The applications will be focused on four-quark mesons with spin $S=0$,
1 or 2, and various flavour content.  As already mentioned, there are
promising possibilities in the exotic sector with two heavy quarks,
especially $(b c \bar{q}\bar{q})$, but these states have not yet been
experimentally searched for.  However, there are indications of
supernumerary states in the charmonium spectrum
\cite{Choi:2003ue,Abe:2004zs,Aubert:2005rm,Swanson:2006st}.  The
single-charm states $(c q \bar{q}\bar{q})$ and analogues with
strangeness were predicted many years ago \cite{Isgur:1980en}, and the
recent findings in the $\mathrm{D}_s$ spectrum might reveal some of
these sates.

The hottest sector is the one of scalar mesons.  Recent experiments at
LEAR and at B-factories have confirmed the years of data taking and
analysis: there are far too many scalar mesons below 2 GeV for the
only $q\bar{q}$ ($q$ denotes $u$ or $d$) and $s\bar{s}$ states, even
including the radial excitations.  The fashion evolved from the
multiquarks of Jaffe to glueballs and hybrids, but seemingly tends
again toward multiquarks.  It is hardly possible to propose an
ultimate solution to this problem.  It appears clearly from the
detailed phenomenological analyses
\cite{Amsler:1995tu,Amsler:1995td,Close:2002zu,Amsler:2004ps,Eidelman:2004wy}
that states with different quark and gluon content are abundantly
mixed and acquire an appreciable mass shift due to their coupling to
the real or virtual decay channels.  Nevertheless, such a mixing
should operate between properly identified bare states, and some
clarification will be suggested in the four-quark sector which is a
key ingredient of the mixing scheme.

This paper is organised as follows: in Sec.~\ref{se:cmh},
the most general chromomagnetic Hamiltonian is presented and
diagonalised
for systems of two quarks and two antiquarks.   The application to
various flavour sectors
is presented in Sec.~\ref{se:app}, before the conclusions in
Sec.~\ref{se:con}.
\section{The chromomagnetic Hamiltonian}\label{se:cmh}
\subsection{General considerations}\label{cmh:sub:gc}
The interaction Hamiltonian acting on the  colour and spin
degrees of freedom, and generalising~(\ref{int:eq:cm1}), is
\begin{equation}\label{cmh:eq:hcm}
H=\sum_{i}m_{i} +H_{\mathrm{CM}}~,\quad
       H_{\mathrm{CM}} = - \sum_{i,j} C_{ij}\, \lala{i}{j}\,
\SpSp{i}{j}~.
\end{equation}
It is inspired by one-gluon-exchange \cite{DeRujula:1975ge}, in which
case $C_{ij}$ contains a factor $\alpha_s/(m_im_j)$, where $\alpha_s$
is the coupling constant of QCD and $m_i$ the mass of the
$i^{\text{th}}$ quark, and also the probability of finding the quarks
(or antiquarks) $i$ and $j$ at the same location.  The above model is
more general. The coefficients $C_{ij}$ which presumably incorporate
non-perturbative QCD contributions, depend on the quark masses and on
the properties of the spatial wave function, as in the
one-gluon-exchange model. The solution of the eigenvalue problem for
the Hamiltonian (\ref{cmh:eq:hcm}) is of interest not only for
spectroscopy, but in all circumstances where two quarks and two
antiquarks are in a relative S-wave, for instance when studying the
violation of the OZI rule \cite{Isgur:2000ts}.

For a quark--antiquark meson, $\langle\lala{1}{2}\rangle=-16/3$ and
$\SpSp{1}{2}=+1$ for spin $S=1$ and $-3$ for $S=0$. The Hamiltonian
(\ref{cmh:eq:hcm}) accounts naturally for the observed hyperfine
splittings such as $\jp-\eta_c$ or $\mathrm{D}_s^*-\mathrm{D}_s$.
This leads to the  strength parameters shown in
Table~\ref{tab:hyperm}.  As the spin-singlet state of  bottomonium
and the
spin-triplet state of $(b\bar{c})$ are not yet known experimentally,
and
in these sectors, the data have been replaced by model
calculations \cite{Eichten:1994gt}.
\begin{table}[!htb]
\caption{\label{tab:hyperm} Values of $C_{q'\bar{q}}$ (in MeV)
estimated from meson masses}
\begin{ruledtabular}
\begin{tabular}{ddddd}
                    & n      &   s    &  c   & b   \\
\hline
\bar{n}  & 29.8                                                  \\
\bar{s}  & 18.4    &  8.6                                     \\
\bar{c}  & 6.6      &  6.7        &    5.5                     \\
\bar{b} & 2.1       & 2.2             & 6.8\footnotemark[1]   &
4.1\footnotemark[1]   \\
\end{tabular}
\end{ruledtabular}
\footnotetext[1]{This is extracted from one of the model calculations
compiled in  Ref.~\protect\cite{Eichten:1994gt}}\end{table}

For ordinary baryons, the colour operator $\lala{i}{j}=-8/3$ is the
same for all pairs and factors out.  For spin $S=3/2$,
$H_{\text{CM}}=8(C_{12}+C_{23}+C_{31})/3$ pushes up $\Delta$,
$\Sigma^*$, etc.  For spin $S=1/2$ $(qqq')$ baryons with two identical
quarks, $H_{\text{CM}}=8/3 (C_{12}-4 C_{13})$ is attractive.  In the
general case $(q_1q_2q_3)$ of spin $1/2$ such as $\Lambda$ or
$\Sigma_0$ with breaking of isospin symmetry, or $\Xi_c^{+}(csu)$, a
basis
\be\label{cmh:eq:bas123}
\left[\left(q_1q_2\right)_1q_3\right]_{1/2}~,\qquad
\left[\left(q_1q_2\right)_0q_3\right]_{1/2}~,
\ee
can be chosen, with symmetric or antisymmetric coupling of the first
two quarks (the index, here and in similar further states, denotes
the value of the spin) in which the chromomagnetic
interaction reads
\be\label{cmh:eq:H123}
H_{\text{CM}}={8\over3}\left[\matrix{C_{12}-2C_{13}-2C_{23} &
\sqrt3(C_{23}-C_{13})\cr
\sqrt3(C_{23}-C_{13}) & -3 C_{12}}\right]~.
\ee

The $\mathrm{N}-\Delta$ system gives access to $C_{qq}$.  Then the
$\Lambda-\Sigma-\Sigma^*$ multiplet gives $C_{qs}$ and another value
of $C_{qq}$ close to the previous one.  Then $\{\Xi,\,\Xi^*\}$ and
$\Omega^-$ depend on $m_s+ 4 C_{ss}/3$ and $m_s+8 C_{ss}/3$ and, to
the extent that these parameters do not change much from $\Xi$ to
$\Omega$, $C_{ss}$ can be obtained.  The value shown for $C_{cc}$ is
from model calculations of double-charm baryons \cite{Fleck:1989mb}.
The values of the strength factors $C_{ij}$ are displayed in Table
\ref{tab:hyperb}.
\begin{table}[!htb]
\caption{\label{tab:hyperb} Approximate values of $C_{ii'}$ (in MeV)
estimated
from baryon masses}
\begin{ruledtabular}
\begin{tabular}{dddd}
                    & n      &   s    &  c     \\
\hline
n  & [19-20]
\\
s  & [12-14]       & [5-10]                                     \\
c  & 4      &  5        &    5\footnotemark[2]
\\
\end{tabular}
\end{ruledtabular}
\footnotetext[2]{This is extracted from one of the model calculations
 in  Ref.~\protect\cite{Fleck:1989mb}}
\end{table}
%


For tetraquarks and higher multiquark states, there is the known
complication that an overall colour singlet can be built from several
manners of arranging internal colour.  These colour states usually can
mix and one has to diagonalise the interaction Hamiltonian.  In the case of
tetraquarks, the most natural basis is constructed by coupling the
quarks $q_1$ and $q_2$ in colour $\bar{3}$ or $6$ and spin $s=0$ or
$1$, to the extent allowed by the Pauli principle, and similarly for
the antiquarks.  However, for studying the decay properties, it is
convenient to translate the state content in the basis
$[(q_1\bar{q}_3)^c(q_2\bar{q})^c]$ or
$[(q_1\bar{q}_4)^c(q_2\bar{q}_3)^c]$.  Here, and in the rest of this
article, the upper index $c$ denotes the colour of the cluster.  It
runs over $c=1$ and $c=8$ in this decomposition.  The relevant
crossing matrices should be derived with care, as some errors and
misprints occurred in the early literature.  In particular, the order
adopted for coupling $q_1$ and $q_2$, for instance, results into phase
factors that do not influence the physics content, but should be
treated consistently throughout the calculation.  The results
presented below have been checked in particular against
\cite{Chan:1980cj} in the limit of isospin symmetry, and
\cite{Wong:1980je}.
\subsection{Group theoretical considerations}
The operator $\mathcal{O}=-\sum \lala{i}{j}\,\SpSp{i}{j}$ can be
elegantly
expressed in terms of the Casimir operators of the
spin \suds,  colour \sutc\ and spin--colour \suscs\ groups,
as stressed in \cite{Jaffe:1976ih,Hogaasen:1978jw,Mulders:1978cp} for
special configurations or more general cases.

For an $N$-constituent system consisting of $n$ quarks and
$\bar{n}=N-n$ antiquarks, with the same strength $C_{ij}=C$ in the
quark sector, $C_{ij}=\overline{C}$ in the antiquark sector, and
$C_{ij}=C'$ for all quark--antiquark pairs, it can be shown that
\begin{eqnarray}\label{ch:eq:sym}
&&2 H_{\text{CM}}=-C\left[C_6(Q)-C_3(Q)-{8\over3} C_2(Q)-16
  n\right]
-\overline{C}\left[C_6(\overline{Q})-C_3(\overline{Q})-{8\over3}
  C_2(\overline{Q})-16 \bar{n}\right]\nonumber\\
&&\quad{}+C'\left[C_6(T)-C_6(Q)-C_6(\overline{Q})
-C_3(T)+C_3(Q)+C_3(\overline{Q}-{8\over3} C_2(T)+{8\over3} C_2(Q)
+{8\over3} C_2(\overline{Q})\right]~,
\end{eqnarray}
where $C_2$, $C_3$ and $C_6$ are the Casimir operators of
SU(2)$_{\text{s}}$, SU(3)$_{\text{c}}$ and SU(6)$_{\text{cs}}$,
respectively, for the quark ($Q$)or antiquark ($\overline{Q}$) sector
or the whole system ($T$). The normalisation adopted here is such that
$C_2=S(S+1)$ for a spin $S$, and $C_3(3)=16/3$
and $C_6(6)=70/3$ for the lowest representations. If it is further
assumed
that $C=\overline{C}=C'$, the well-known formula \cite{Jaffe:1976ih}
\begin{equation}\label{ch:eq:sym1}
\mathcal{O}=8 N +{1\over2} C_6(T)-{4\over3}C_2(T)-{1\over2}C_3(T)+
C_3(Q)+{8\over3}C_2(Q)-C_6(Q)+C_3(\overline{Q})
+{8\over3}C_2(\overline{Q})-C_6(\overline{Q})~,
\end{equation}
is recovered.

It is possible to make some general considerations on the eigenvalues
of the chromomagnetic interaction for the scalar, axial and tensor
tetraquarks.  Consider first the flavour-symmetry limit, which is a
good approximation for the states built from light ($q=u,\,d$) quarks
and antiquarks.  In this limit, the matrix representation
$H_{\text{CM}}$ simplifies to two $2\times2$ matrices for the scalars,
two $2\times2$ and two $1\times1$ for the axials and two $1\times1$
for the tensors.  The interaction between the quarks and the
antiquarks, which depends strongly on the SU(6)$_{\text{cs}}$ Casimir
operators of the tetraquark, has a tendency to give eigenstates which
approximately belong to the irreducible representations of that
algebra.

This observation has also interesting consequences for the decay
properties of tetraquarks.  In fact, many years ago, Jaffe
\cite{Jaffe:1976ig,Jaffe:1976ih} stressed that all the multiquarks
have ``open door'' channels, that is to say, can decay into two colour
singlets by simple rearrangement of the constituents,  see, also,
Refs.~\cite{Petry:1983,Bleuler:1984}.  Only phase space can possibly
block this spontaneous dissociation.

More recently, this property has been related \cite{Buccella:2004xd}
to the transformation properties of the multiquark states with respect
to \suscs. Since the pseudoscalar ($\pi$, K, $\eta$, $\eta'$) and the
vector ($\rho$, K$^*$, $\omega$, $\phi$) mesons transform as a singlet
and a 35, respectively, the ``open door'' pseudoscalar--pseudoscalar
(PP) channels will be \suscs\ singlets and the pseudoscalar--vector
(PV) channels will be 35-plets of the same algebra. The `open door''
vector--vector (VV) channels will be found for the states transforming
in a representation contained in the product of two 35 representations
(1, 35, 189, 280, $\overline{280}$ and 405).

The scalar states built from light quarks belong to the
representations $1+405$ of \suscs\ for the case of isospin $I=0$ and
to the representations $1+189$ for $I=0,1,2$. Indeed, the quarks
symmetric (resp.\ antisymmetric) in colour--spin belong to the
$(6\times 6)_{\text{S}}=21$ (resp.\ $(6\times 6)_{\text{A}}=15$
representations of \suscs. From the decomposition of the \suscs\
representations with respect to $\sutc\times\suds$
\begin{equation}
21=(6,3)+(\overline{3},1)~,\qquad
15=(6,1)+(\overline{3},3)~,
\end{equation}
and the \suscs\ products of representations
\begin{equation}
21\times\overline{21}=1+35+405~,\qquad
15\times\overline{15}=1+35+189~.
\end{equation}
it is readily seen that two $(1,1)$ singlets of
$\text{SU(3)}_{\text{c}}\times\text{SU(2)}_{\text{s}}$
come from the $21\times\overline{21}$ and $15\times \overline{15}$
products, and also that the $35$ representation does not contain any 
$(1,1)$ singlet of
$\text{SU(3)}_{\text{c}}\times\text{SU(2)}_{\text{s}}$.

In order to apply Eqs.~(\ref{ch:eq:sym1}) and (\ref{ch:eq:sym}) to
these
states,  the following SU(6) Clebsch--Gordan coefficients are
necessary
\begin{eqnarray}\label{ch:eq:6times6a}
\left|1\right\rangle&=&\sqrt{6\over7}\left|
21;(6,3)\right\rangle\left|
\overline{21};(\overline{6},3)\right\rangle
+{1\over\sqrt7}\left| 21;(\overline{3},1)\right\rangle\left|
\overline{21};(3,1)\right\rangle~,\nonumber\\
\left |405 \right\rangle&=&{1\over\sqrt7}\left| 21;(6,3)\right\rangle
\left|
\overline{21};(\overline{6},3)\right\rangle-\sqrt{6\over7}\left|
21;(\overline{3},1)\right\rangle\left|
\overline{21};(3,1)\right\rangle~,\nonumber\\
\left|1\right\rangle&=&\sqrt{2\over5}\left|
15;(6,1)\right\rangle\left|
\overline{15};(\overline{6},1)\right\rangle+\sqrt{3\over5}\left|
15;(\overline{3},3)\right\rangle\left|
\overline{15};(3,1)\right\rangle~,\nonumber\\
\left |189 \right\rangle&=&\sqrt{3\over5}\left|
15;(6,1)\right\rangle\left|
\overline{15};(\overline{6},1)\right\rangle
-\sqrt{2\over5}\left| 15;(\overline{3},3)\right\rangle\left|
\overline{15};(3,1)\right\rangle~.
\end{eqnarray}

Now, the \suscs\ Casimir dependence of the chromomagnetic contribution
to the mass of the tetraquarks shown in Eq.~(\ref{ch:eq:6times6a})
implies that the lightest states are approximately singlets, while the
heavier states transforming approximately as the 405 or the 189
representation, have large coupling to VV and small coupling to PP
channels.

As for the axial sector, the lightest state will be a $I=0$
transforming as a 35, followed by two $I=1$ states and a $I=0,\,1,\,2$
cluster transforming in the same way, while the heaviest states are
the two $I=1$ transforming approximately as $280+\overline{280}$.  Due
to parity conservation, a $1^+$ state cannot decay into two
pseudoscalar mesons, the heaviest states are expected to have a small
amplitude to PV and may lie below the threshold for VV. Note that the
four 35 may be too light to decay into PV.

Finally, the tensor states, which have S-wave amplitudes into VV, may
be under threshold for that final state.

When states with one or more strange constituents are considered, the
chromomagnetic interaction involve different gyromagnetic factors and
short-range correlations.  These symmetry-breaking effects mix states
with different \suscs\ transformation properties for the $qq$ and
$\bar{q}\bar{q}$ pairs, but many of the qualitative features of the
symmetry limit remain, both for the hierarchy of masses and decay
patterns.  However, for detailed phenomenological applications, it is
desirable to have explicit estimates of the eigenstates of
$H_{\text{CM}}$, and for this purpose, instead of using a basis of
\suds, \sutc\ and \suscs\ representations, it is preferable to couple
explicitly the quarks in states of given spin and colour, and
similarly for the antiquarks.  This new basis turns out also more
convenient to impose the constraints due to Pauli principle.  The
calculations are now carried out in some detail for the scalar, axial
and tensor configurations.

\subsection{Scalar tetraquarks}\label{cmh:sub:sca}
Consider first the case of total spin $S= 0$.  In the
$[(q_1q_2)(\bar{q}_3\bar{q}_4)]$ basis, the diquark and the
antidiquark should bear conjugate colour, $(\bar{3},3)$ or
$(6,\bar{6})$, and the same spin $0$ or $1$.  The Hamiltonian
(\ref{int:eq:cm1}) acts on the four states:
\begin{eqnarray}\label{cmh:eq:bas0}
& \phi_1 =(q_1 q_2)^6_1  \otimes  (\bar{q}_3 \bar{q}_4
)^{\bar{6}}_{1}~,\quad
&\phi_2 =(q_1 q_2)^{\overline{3}}_0 \otimes  (\bar{q}_3
\bar{q}_4)^{3}_{0}~,\nonumber \\
& \phi_3 =(q_1 q_2)^6_0  \otimes  (\bar{q}_3
\bar{q}_4)^{\bar{6}}_{0}~,  \quad
&\phi_4 = (q_1 q_2)^{\overline{3}}_1 \otimes  (\bar{q}_3
\bar{q}_4)^{3}_{1}~.
\end{eqnarray}
The colour-magnetic interaction in this basis reads
\begin{equation}\label{cmh:eq:scaH1}
       H_{\text{CM} }= - \left[ \matrix{A_1 &A_2\cr
B_1 &B_2} \right],
\end{equation}
with $2 \times 2$ submatrices
\begin{widetext}
\begin{eqnarray}
\label{cmh:eq:scaH2}
A_1&=&\left[ \matrix{%
\df{4}{3}(C_{34}+C_{12})+\df{20}{3}(C_{14}+C_{13}+C_{23}+C_{24})&
2\sqrt6\left( C_{14}+C_{13}+C_{23}+C_{24}\right)\cr
2\sqrt6\left( C_{14}+C_{13}+C_{23}+C_{24}\right)&
8(C_{34}+C_{12})}\right]~,\nonumber\\[2pt]
A_2=B_1^\dagger&=&{2\over\sqrt{3}}\,\left( C_{13}-C_{14}+
C_{24}-C_{23}
\right) \,\left[ %
\matrix{%
5& 2 \sqrt{6}\cr
0&2               }                     \right]~,\\[2pt]
B_2 &=&  \left[ \matrix{%
 -4\,(C_{34}+C_{12})&
 2 \sqrt {6} \left(C_{14}+C_{13}+C_{23}+C_{24} \right)\cr
2\sqrt {6}\left( C_{14}+C_{13}+C_{23}+C_{24} \right) &
 -\df{8}{3}\,(C_{34}+C_{12}-C_{14}-C_{13}-C_{23}-C_{24})
 }\right]~.\nonumber
\end{eqnarray}
\end{widetext}
In the states $\phi_1$ and $\phi_2$, the quarks are symmetric in
colour--spin and belong to the $(6\times 6)_{\text{S}}= 21$
dimensional representation of SU(6)$_{\text{cs}}$, and the antiquarks
belong to a $\overline{21}$ representation.  In $\phi_3$ and $\phi_4$,
the quarks are coupled antisymmetrically in a $(6 \times 6)_{\text{A}
}= 15$ representation, and the antiquarks in a $\overline{15}$.  If
only three flavours are involved, $\phi_1$ and $\phi_2$ fall into the
$\bar{3}\times 3=1+8$ representations of SU(3)$_{\text{F}}$, which is
called a nonet in the familiar notation of this symmetry group, and
$\phi_3$ and $\phi_4$ fall into the $6\times \bar{6}=1+8+27$
representations.

If the two quarks $q_{1}$ and $q_{2}$ or the two antiquarks
$\bar{q}_{3}$ and $\bar{q}_4$, are identical, the states $\phi_1$ and
$\phi_2$ are excluded by the Pauli principle, and in the space spanned
by $\phi_3$ and $\phi_4$, the Hamiltonian $H_{\text{CM}}$ is expressed
by the $2\times 2$ matrix $-B_2$.

In the limit where one antiquark, say $\bar{q}_4$, is very heavy and
decouples, i.e., $C_{i4}$=$C_{4i}=0$, the problem reduces to the
previously discussed \cite{Hogaasen:2004pm,Hogaasen:2004ij}
chromomagnetic problem of a spin 1/2, colour triplet $(qq\bar{q})$
triquark.  It always contains a colour singlet $q\bar{q}$ pair,
leading to superallowed decays, if kinematically permitted.

In the flavour-symmetry limit, with the further assumption that the
quark--quark and quark--antiquark colour--spin interaction strengths
are equal, $H_{\text{CM}}$ reduces to
\begin{equation}\label{cmh:eq:su3}
H_{\mathrm{CM}} = -C  \left[ \matrix{%
{88}/{3}&8\sqrt {6}&0&0\cr
8\sqrt {6}&16&0&0\cr
0&0&-8&8\sqrt {6}\cr
0&0&8\sqrt {6}&{16}/{3}
}\right],
\end{equation}
with eigenvalues $-43.3656C$ and $-1.9678C$ in the nonet subspace
spanned by $\phi_1$ and $\phi_2$, and $ -19.3656C$ and $+22.0322C$ in
the 36-plet spanned by $\phi_3$ and $\phi_4$, which separates out
exactly.  The lightest state in the nonet and the lightest one in the
36-plet are split by $24C$, i.e., about 400 MeV, exceeding twice the
mass difference between strange and non-strange quarks.  This led one
to predict that the flavour nonet and 36-plet are well separated.  It
will be shown later that this is not any longer the case, if flavour
symmetry is broken also in $H_{\text{CM}}$ (and not only in the
constituent masses).

From Eqs.~(\ref{cmh:eq:scaH1}-\ref{cmh:eq:scaH2}), it is seen that for
the separation of the 36-plet from the nonet to remain, with a
block-diagonal form for $H_{\mathrm{CM}}$, it suffices that
$C_{13}=C_{14}$ and $C_{23}=C_{24}$, or $C_{13}=C_{23}$ and
$C_{14}=C_{24}$, i.e., both quarks have the same coupling to each
antiquarks, or vice-versa.  It also persists that the lowest
eigenvalue is found in the nonet subspace spanned by $\phi_1$ and
$\phi_2$.

\begin{figure}[!ht]
\begin{center}
\includegraphics[width=.5\textwidth]{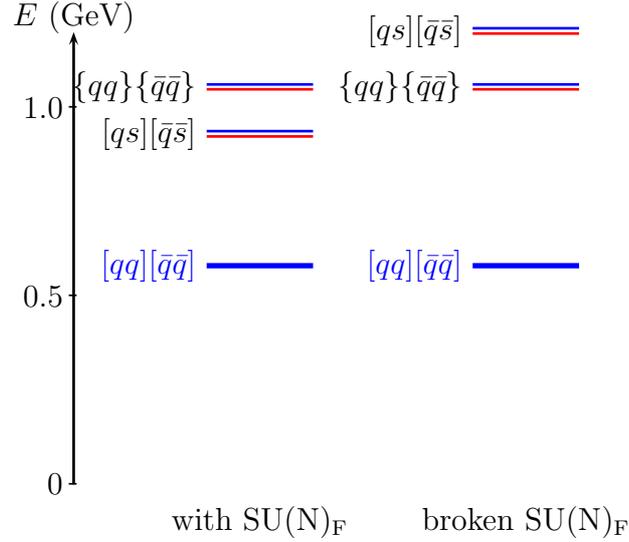}
\end{center}
\caption{\label{Fig1} Spectrum of light multiquark scalars, with a
shift $m_s-m_q$ added for each strange quark or antiquark, and the
chromomagnetic term calculated either in the SU(N)$_\mathrm{F}$ limit
(left) or with SU(N)$_\mathrm{F}$ breaking (right).  In the labels,
$[qq]$ denotes the symmetric spin-colour coupling, and $\{qq\}$ the
antisymmetric one, and $q=u,\,d$.}
\end{figure}

As an illustration, the mass spectrum of light $0^+$ mesons is shown
in Fig.~\ref{Fig1} with and without flavour symmetry breaking in
$H_{\text{CM}}$, with realistic values for the strength factors
$C_{ij}$.

For completeness, the crossing matrix is provided between the basis
(\ref{cmh:eq:bas0}) where quarks and antiquarks are paired, and the
basis
\begin{eqnarray}\label{cmh:eq:bas1}
&\alpha_1 =(q_1 \bar{q}_3)^1_0  \otimes (q_2 \bar{q}_4
)^1_{0}~,\quad
&\alpha_2 =(q_1 \bar{q}_3)^1_1 \otimes (q_2 \bar{q}_4
)^1_{1}~,\nonumber\\
& \alpha_3 =(q_1 \bar{q}_3)^8_0  \otimes  (q_2
\bar{q}_4 )^8_{0}~,\quad
&\alpha_4 =(q_1 \bar{q}_3)^8_1  \otimes  (q_2 \bar{q}_4 )^8_{1} ~,
\end{eqnarray}
with quark--antiquark coupling, it  is
\begin{equation}\label{cmh:eq:cross1}
{1\over 6}\left[ \matrix{%
3\sqrt{2}&\sqrt{3}&\sqrt{6}&3\cr
-\sqrt{6}&3&3\sqrt{2}&-\sqrt{3}\cr
3&-\sqrt{6}&\sqrt{3}&-3\sqrt{2}\cr
-\sqrt{3}&-3\sqrt{2}&3&\sqrt{6}
}\right]~.
\end{equation}
\subsection{Axial tetraquarks}
The case where the total spin is $S=1$ is somewhat more complicated
 than the spin $S=0$ case
 as the recoupling to spin 1 can be done in several ways.
 The colour-magnetic Hamiltonian now acts over a six-dimensional space
with basis

\begin{eqnarray}\label{cmh:eq:bas2}
& \psi_1 =(q_1 q_2)^6_1\otimes(\bar{q}_3 \bar{q}_4
)^{\bar{6}}_{1}~,\quad
& \psi_2 =(q_1 q_2)^{\bar{3}}_1\otimes(\bar{q}_3 \bar{q}_4)^3_{1}~,
\nonumber \\
& \psi_3 =(q_1 q_2)^{\bar{3}}_0\otimes(\bar{q}_3
\bar{q}_4)^3_{1}~,\quad
&\psi_4 =(q_1 q_2)^6_1\otimes(\bar{q}_3 \bar{q}_4)^{\bar{6}}_{0}~,\\
&\psi_5 =(q_1 q_2)^{\bar{3}}_1\otimes(\bar{q}_3
\bar{q}_4)^3_{0}~,\quad
&\psi_6 =(q_1 q_2)^6_0\otimes(\bar{q}_3
\bar{q}_4)^{\bar{6}}_{1}~.\nonumber
\end{eqnarray}
If the two quarks (antiquarks) are identical in flavour,
 $\psi_1$,  $\psi_3$ and $\psi_4$  ($\psi_1$,  $\psi_5$ and $\psi_6$)
are excluded by the Pauli principle. This is the case in particular
for the manifestly exotic states.

The colour-magnetic Hamiltonian can be  written in terms of $2\times2$
blocks as
\begin{equation}\label{cmh:eq:axABdef}
       H_{\text{CM}} = - \left[\matrix{A_1&A_2&A_3\cr  B_1 &
B_2&B_3\cr
C_1 & C_2&C_3} \right],
\end{equation}
with
\begin{eqnarray}\label{cmh:eq:axABi}
 A_1 &=&{2\over 3}\left[ \begin {array}{cc}
2(C_{34}+C_{12})+5(C_{14}+C_{13}+C_{23}+C_{24})&
3\sqrt{2}\left( C_{13}-C_{14}+C_{24}-C_{23} \right)\\
3\sqrt{2}\left( C_{13}-C_{14}+C_{24}-C_{23} \right)&
\!\!\! -4(C_{34}+C_{12})+2\left(C_{14}+C_{13}+C_{23}+C_{24}\right)\\
\end{array} \right]~,\nonumber\\
 A_2=B_1^\dagger&=& {2\over3}\left[ \begin {array}{cc}
6\left(C_{13}-C_{14}+C_{23}-C_{24}\right)&
-5\sqrt {2} \left( C_{13}-C_{14 }+C_{23}-C_{24} \right) \\
2\sqrt {2} \left( C_{13}+C_{14}-C_{24}-C_{23} \right)&
-6\left(C_{13}+C_{14}-C_{24} -C_{23}\right)
 \end {array}\right]~,\nonumber\\
A_3 = C_1^\dagger&=&{2\over3} \left[ \begin {array}{cc}
-6\left(C_{13}+C_{14}-C_{24} -C_{23}\right)&
5\sqrt {2} \left( C_{13}+C_{14}-C_{24}-C_{23} \right)\\
-2\sqrt {2} \left( C_{13}-C_{14}+C_{23}-C_{24} \right) &
6\left(C_{13}-C_{14}+C_{23}-C_{24}\right)
\end {array} \right]~,\nonumber\\
 B_2&=&{2\over3} \left[ \begin {array}{cc}
 4\left(3C_{12}-C_{34}\right) &  -3\sqrt {2} \left( C_{13}+C_{14
}+C_{23}+C_{24} \right) \\
 -3\sqrt {2} \left( C_{13}+C_{14 }+C_{23}+C_{24} \right) &
2C_{12}-6C_{34}
\end {array} \right]~,\\
B_3 = C_2^\dagger&=&{2\over3} \left[ \begin {array}{cc}
 -2\left(C_{13}-C_{14}+C_{24}-C_{23}\right)&0 \\
 0 & -5\left(C_{13}-C_{14}+C_{24}-C_{23}\right)
 \end {array} \right]~,\nonumber\\
 C_3&=& {2\over3  }\left[ \begin {array}{cc}
 -4C_{12}+12C_{34}&-3\sqrt{2}\left( C_{14}+C_{13}+C_{23}+C_{24}
\right)\\
 -3\sqrt{2}\left(C_{14}+C_{13}+C_{23}+C_{24} \right) &-6C_{12}+2C_{34}
\end {array} \right]~.\nonumber
\end{eqnarray}
In the limit of flavour symmetry where $C_{ij}=C$, $\forall i,j$, the
eigenstates of $H_{\text{CM}}$ have well defined transformation
properties under the relevant flavour-symmetry group, and the
colour-magnetic Hamiltonian $H_{\text{CM}}$ reduces to the well-known
matrix
\begin{equation}\label{cmh:eq:axfs}
-{8C\over 3}  \left[\matrix{%
6&0&0&0&0&0\cr
0&0&0&0&0&0\cr
0&0&2&-3\sqrt{2}&0&0\cr
0 &0&-3\sqrt{2}&-1&0&0\cr
0&0&0&0&2&-3\sqrt {2}\cr
0&0&0&0&-3\sqrt {2}&-1} \right]~.
\end{equation}
 with eigenvalues $-16C$, $0$, $-40C/3$, $32C/3$, $-40C/3$, $32C/3$.
 The corresponding flavour multiplets are $9$ and $36$ for the first
 eigenvalues, $\overline{18}=\overline{10}+8$ for the next two ones,
 and $18=10+8$ for the last two ones

 Moreover, for the interesting case $(QQ\overline{u}\overline{d})$
 case where the two heavy quarks $Q$ are identical and the two light
antiquarks obey isospin symmetry, $H_{\text{CM}}$ also takes the
block-diagonal form
\begin{equation}\label{cmh:eq:axQQ}
-{4\over3}\left[ \begin {array}{cccccc}
C_{34}+C_{12}+10\,C_{14}&0&0&0&0&0\\
0&\!\! -2\,C_{34}-2\,C_{12}+4\,C_{14}&0&0&0&0\\
0&0&\!\!-2\,C_{34}+6\,C_{12}&-6\sqrt{2}\,C_{14}&0&0\\
0&0& -6\sqrt {2}\,C_{14}&C_{12}-3\,C_{34}&0&0\\
0&0&0&0&\!\!-2\,C_{12}+6\,C_{34}&-6\sqrt {2}\,C_{14}\\
0&0&0&0&\!\!-6\sqrt{2}\,C_{14}&-3\,C_{12}+C_{34}
\end {array} \right]
\end{equation}

Note that, contrary to what happens for the spin $S=0$ case, the
lowest eigenvalue of the colourmagnetic Hamiltonian survives the Pauli
principle, i.e., remains when the basis states $\psi_1$, $\psi_3$ and
$\psi_4$ are removed, at least for all the physically acceptable
values of the parameters (see next section).

The crossing matrix from the basis (\ref{cmh:eq:bas2}) to the basis
\begin{eqnarray}\label{cmh:eq:bas3}
&  \beta_1 =(q_1 \bar{q}_3)^1_0 \otimes(q_2 \bar{q}_4
)^{{1}}_{1}~,\quad
&\beta_2 =(q_1 \bar{q}_3)^1_1\otimes (q_2 \bar{q}_4
)^{{1}}_{0}~,\nonumber \\
&  \beta_3 =(q_1 \bar{q}_3)^1_1\otimes (q_2 \bar{q}_4
)^{{1}}_{1}~, \quad
&\beta_4 =(q_1 \bar{q}_3)^8_0\otimes (q_2 \bar{q}_4 )^{{8}}_{1}~, \\
&  \beta_5 =(q_1 \bar{q}_3)^8_1\otimes(q_2 \bar{q}_4
)^{{8}}_{0}~, \quad
&\beta_6 =(q_1 \bar{q}_3)^8_1\otimes(q_2 \bar{q}_4
)^{{8}}_{1}~,\nonumber
\end{eqnarray}
is
\begin{equation}\label{cmh:eq:cross2}
{1\over 2\sqrt3}\left[\matrix{
2&\sqrt2&1&-\sqrt2&-1&\sqrt2\cr
2&\sqrt2&-1&\sqrt2&1&-\sqrt2\cr
0&0&\sqrt2&2&\sqrt2&2\cr
\sqrt2&-2&-\sqrt2&-1&\sqrt2&1\cr
\sqrt2&-2&\sqrt2&1&-\sqrt2&-1\cr
0&0&-2&\sqrt2&-2&\sqrt2
}\right].
\end{equation}

\subsection{Tensor tetraquarks}
The survey is ended by the case of spin $S=2$. In the
diquark--antidiquark coupling scheme, the chromomagnetic
Hamiltonian $H_{\text{CM}}$, written in the basis
\begin{equation}\label{cmh:eq:bas4}
\xi_1=(q_1 q_2)^{6}_1\otimes (\bar{q}_3
\bar{q}_4)^{\bar{6}}_{1}~,\quad
 \xi_2=(q_1 q_2)^{\bar{3}}_1\otimes (\bar{q}_3
\bar{q}_4)^{\bar{3}}_{1}~,
 \end{equation}
reads
\begin{equation}\label{cmh:eq:tens1}
-  \frac{2}{3}\left[ \begin {array}{cc}
2\left(C_{12}+C_{34}\right)-5\left(C_{13}
+C_{24}+C_{14}+C_{23}\right) &
 - 3 \sqrt {2} \left(  C_{13}+C_{24}- C_{23}- C_{14} \right)\\
 - 3 \sqrt {2} \left( C_{_{13}}+C_{24}-C_{23}-C_{14}\right) &
-4\left (C_{12}+C_{34}\right)-2\left(C_{13}+C_{24} +
C_{23}+C_{14}\right)
\end {array}\right]~.
\end{equation}
With two quarks identical in flavour, the state $\xi_1$ is excluded
by the Pauli principle.

As all spins are aligned, the crossing matrix from the basis
(\ref{cmh:eq:tens1}) to the basis
\begin{equation}\label{cmh:eq:bas5}
\gamma_1=(q_1 \bar{q}_3)^{1}_1\otimes
({q}_2\bar{q}_4)^{\bar{1}}_{1}~,\quad
\gamma_2=(q_1 \bar{q}_3)^{8}_1\otimes ({q}_2\bar{q}_4)^{\bar{8}}_{1}~,
 \end{equation}
 reduces to the standard crossing matrix of colour
\begin{equation}\label{cmh:eq:cross3}
{1\over \sqrt3}\left[\matrix{
\sqrt2&1\cr
1&-\sqrt 2}\right]~.
\end{equation}
\section{Application to tetraquarks}\label{se:app}
This section is devoted to consequences of the chromomagnetic
interaction
applied to four-quark states for the various flavour configurations. 
\subsection{Adjusting the parameters}
The strength parameters $C_{ij}$ for quark--antiquark pairs can be
extracted from ordinary mesons, and are given in
Table~\ref{tab:hyperm}. They can be considered as upper bounds, as
the two-body correlations are stronger in mesons than in tetraquarks.
The quark--quark analogues, deduced from the baryon spectrum, are
listed in Table~\ref{tab:hyperb}.
These parameters can be used to extrapolate the model from ordinary
hadrons to multiquarks.

A tempting alternative strategy consists of extracting the parameters
from states which are assumed to be dominantly tetraquarks
\cite{Chen:2006hy} and to apply the model to predict new tetraquark
states.  However, the observed states very likely result from an
intricate mixing of four-quark, two-quark, hybrid and gluonium states,
and the fit can be biased if this mixing is ignored.

There is no way to determine the effective masses unambiguously, as
they incorporate binding effects which depend on the environment.  In
particular, the values of $m_i$ extracted from baryons are usually
higher than those from mesons.  This is, indeed, a general property
that baryons are heavier, per quark, than mesons, for instance
$\mathcal{M}(\Omega)/3> \mathcal{M}(\phi)/2$.  The inequality
$(qqq)/3> (q\bar{q})/2$ can be derived in a large class of models
inspired by QCD~\cite{Nussinov:1999sx}; in this review article, and
refs.\ therein, it is also reminded that $(q\bar{q})+(Q\bar{Q})\le 2
(Q\bar{q})$, hence masses deduced from hidden flavour are found
lighter than from open flavour.  In a multiquark such as
$(c\bar{c}q\bar{q})$, a compromise has be found, as in
Ref.~\cite{Hogaasen:2005jv}.

The chromomagnetic model will thus be used mainly to predict the
ordering of the various spin and flavour configurations.  Estimating
the absolute masses would require a more careful treatment of the
chromoelectric effects.
\subsection{Light mesons}\label{sub:light}
\begin{figure}[ht!]
\begin{center}
\includegraphics[width=.5\textwidth]{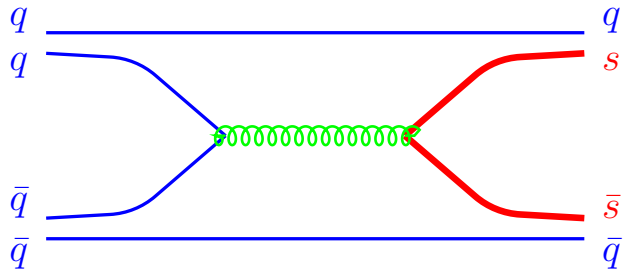}
\end{center}
\caption{\label{Fig2} Example of mixing through internal annihilation}
\end{figure}

This is the most delicate sector.  Experimentally, states are often
broad and overlapping.  Theoretically, the quark--antiquark spectrum
is not as easily described as in the case of heavy quarks, and states
with exotic internal structure are thus harder to single out.
Moreover, mixing of configurations is more appreciable in this sector.
Just to illustrate the complexity, the diagrams with internal $q
\bar{q}\leftrightarrow s\bar{s}$ transition through an intermediate
gluon (see Fig.~\ref{Fig2}) mix \emph{ten} tetraquark states with
$I=0$, and \emph{six} with $I=1$ \cite{Chan:1980cj}.  Hence great care
is required when discussing this sector.

There are many scalar mesons below about $2\;$GeV, and different
scenarios have been proposed for their assignment, see, e.g.,
\cite{Amsler:2004ps,Eidelman:2004wy} and references therein.  For
instance, Klempt et al.\ \cite{Klempt:1995ku}, using a relativistic
quark model and an instanton-induced interaction, proposed that the
mainly quark--antiquark multiplet includes $\mathrm{f}^0(980)$ and
$\mathrm{f}^0(1500)$.  Alternatively, it is tempting to assign
$\mathrm{f}^0(1370)$ as being mainly $q\bar{q}$, in its
${}^3\mathrm{P}_0$ configuration, which is expected to lie slightly
below its ${}^3\mathrm{P}_1$ (perhaps mixed with ${}^1\mathrm{P}_1$)
and ${}^3\mathrm{P}_2$ partners, in analogy with what is observed in
the case of charmonium.  This is the point of view adopted, e.g., in
\cite{Amsler:1995td,Amsler:1995tu}, where the $\mathrm{f}^0(1500)$ is
tentatively identified with the lowest gluonium state, the second
$q\bar{q}$ state with isospin $I=0$ being slightly higher.  The expert
view point of the latest issue of the review of particles properties
\cite{Eidelman:2004wy} suggests to organize the scalars in a low-lying
nonet consisting of $\mathrm{f}^0(600)$, $\mathrm{f}^0(980)$,
$\mathrm{a}(980)$ and $\mathrm{K}_0^*(800)= \kappa$, a second
multiplet being made of $\mathrm{f}^0(1370)$, $\mathrm{f}^0(1500)$,
$\mathrm{a}(1450)$ and $\mathrm{K}_0^*(1430)$, with the caveat that
the $\mathrm{f}^0(1500)$ is copiously mixed with the
$\mathrm{f}^0(1710)$ to share their $q\bar{q}$, $s\bar{s}$ and
gluonium content.  This is not too far from the recent analysis by
Narison~\cite{Narison:2005wc}.

If it is assumed that $C_{qq}=C_{q\bar{q}}$,
$C_{qs}=C_{q\bar{s}}=0.625 C_{qq}$, and
$C_{ss}=C_{s\bar{s}}=C_{qs}^2/C_{qq}$, and if the values
$C_{qq}=19.2\;$MeV, $m_q=320\;$MeV and $m_s=445\;$MeV are adopted, the
two above nonets come with masses (439, 722, 980)\,MeV and (1242,
1376, 1512)\,MeV, respectively, in our simple chromomagnetic model.
The agreement with experimental masses is perhaps too good, as mixing
with other configurations is expected to shift these results.  Anyhow,
these parameters also give the remaining tetraquark spectrum.  In
particular, for the $Y=2$ axials considered in \cite{Chen:2006hy} with
$(qq\bar{s}\bar{s})$ content, a first state is found at $1310\;$MeV
and a heavier one at $1620\;$MeV, while for the 27 ($I=1$) $Y=2$, the
mass is predicted to be about $1540\;$MeV. States with both open and
hidden strangeness are obtained at 1510 and 1870 Mev ($0^+$), 1500,
1640 and 1760 MeV ($1^+$) and $1810\;$ MeV ($2^+$), and an axial state
with a double hidden strangeness is predicted at $1780\;$MeV.
Moreover, the $\phi\omega$ resonance found at BES II
\cite{Ablikim:2006dw} at $1812\;$MeV could be identified with the
multiquark scalar with hidden strangeness.

A difficulty with this description of tetraquarks, or at least, with 
this choice of parameters, is the prediction of a scalar multiplet
with $I=0,\,1,\,2$ at about $800\;$MeV, without experimental evidence
for a $I=1$ resonance nor for a $I=2$ one in that
region.\footnote{Note that the repulsive character of the $I=2$
$\pi\pi$ phase-shift in this region is not a definite obstacle, as
well the pentaquark is not ruled out solely by the negative values of
the $\mathrm{KN}$ phase-shift.  If one imagines a hadron--hadron
potential with a repulsive long-range tail and an attractive
short-range well due to quark dynamics, low-energy scattering
experiments will hardly detect the presence of the inner hole and will
only feel the repulsive tail.} Also, a puzzlingly light $I=0$ state
would be predicted with this set of parameters.

An alternative strategy consists of adopting the parameters $m_i$ and
$C_{ij}$ deduced from
baryon masses. With this choice of parameters,
the main message is conveyed by Fig.\ \ref{Fig1}.
There is an isolated isoscalar which can be identified with $\sigma$.
Next
comes a nearly degenerate set of $I=0$, $I=1$ and $I=2$ states
corresponding to $\mathrm{f}^0(980)$, $\mathrm{a}(980)$ and a yet not
discovered $I=2$ state. The states with one unit of strangeness
correspond
to $\kappa$. The first state with hidden strangeness is at about 1050
MeV.

Note that with $C_{qs}<C_{qq}$ , which is expected as a consequence of
flavour symmetry breaking, the states with hidden strangeness will be
pushed up in mass relative to their mass in the flavour symmetric
case.  Indeed as can be seen from Fig.~\ref{Fig1},with that choice of
parameters, the states with hidden strangeness ($I=0$ and $I=1$) in
the flavour nonet is more massive than the $I=0,\,1$ and 2 states
without hidden strangeness in the flavour 36-plet!  A bag model
calculation found these nonet and 36-plet states at almost the same
mass and the authors considered their mixing with interesting results
\cite{Bickerstaff:1982uk}%
\footnote{To us it seems that this mixing could be more important than
the mixing between $0^+$ states inside the flavour nonet estimated by
F.~Giacosa et.  al.\ \protect\cite{Giacosa:2006rg}.}.  This is at
significant variance with the results of \cite{Jaffe:1976ig}, which
inspired several phenomenological analyses.  In short, the
$\mathrm{a}_0$ and $\mathrm{f}_0$, if described as multiquark hadrons,
do no acquire much hidden strangeness from short-range, direct quark
interaction.  Indeed, the chromomagnetic effects are significantly
weaker for strange quarks than for ordinary quarks.  The mixing is
particularly important for light scalar mesons with $I=0$ and, to a
lesser extent $I=1$, because the wave function obtained by
diagonalisation of $H_{\text{CM}}$ contains large components where a
quark-antiquark pair has spin triplet and colour octet.

The above discussion suggests that it is diffcult to explain the
low-lying $0^+$ mesons as tetraquark states without mixing to other
configurations, even if flavour-symmetry breaking is treated
consistently in the tetraquark sector.  One should note the inverted
nonet of light scalar mesons, that comes out naturally from the
chromomagnetic interaction in light tetraquarks, also arises in a
multichannel picture where colour-singlet quark--antiquark states are
coupled to meson--meson channels
\cite{VanBeveren:1986ea,vanBeveren:2006ua,Tornqvist:1995kr}.
This is also the conclusion of Maiani et al.\ \cite{Maiani:2006rq} that light tetraquark scalars lie below the quark--antiquark ones.

\subsection{Mesons with heavy flavour}
This sector has been particularly discussed recently, following the
observation by several groups of the $\mathrm{D}_s(2317)$ and
$\mathrm{D}_s(2457)$ resonances with $(c\bar{s})$ flavour content
\cite{Eidelman:2004wy}.  The current wisdom is that these states have
$J^P=0^+$ and $1^+$ quantum numbers, respectively.  In standard quark
models, see, e.g., \cite{Cahn:2003cw}, the masses of the
${}^3\mathrm{P}_0$ and ${}^3\mathrm{P}_1$ (perhaps mixed with
${}^1\mathrm{P}_1$) ground states are, by about 100\,MeV too high to
match the observed masses, and the $\mathrm{D}_{s,J}$ states are
therefore to be described with other tools.  See, however,
\cite{Deandrea:2003gb,Matsuki:2006pt}, where the new states with open
charm are accommodated without a need for any exotic structure.

The fashionable ``molecular picture'' has been proposed for these
states, and was even anticipated before their discovery, see, e.g.,
\cite{Barnes:2003dj,Bicudo:2004dx} and references there.  In another
scenario, these states are described as chiral partners of the
pseudoscalar $\mathrm{D}_s( 1968)$ and vector $\mathrm{D}_s( 2112)$
ground states.  Again, the restoration of chiral symmetry in this
sector was predicted as reminded in the recent analysis
\cite{Bardeen:2003kt} where references are given to the earlier
papers.  The doubling of $\mathrm{D}_s$ states by chiral symmetry, is,
however, not universally accepted, as discussed, e.g.,
\cite{Jaffe:2005sq}.

If the $\mathrm{D}_s(2317)$ and $\mathrm{D}_s(2457)$ are basically
four-quark states, as suggested, e.g., in
\cite{Lipkin:2003zk,Terasaki:2005ne}, then the states with
dominant quark--antiquark content bearing the same quantum numbers
await experimental discovery.

In the calculation based on our chromomagnetic Hamiltonian and our
preferred choice of parameters, the $0^+$ and $1^+$ multiquark states
come too high as compared to the experimental masses and are thus
rather broad.  Still, they produce a mixing with the $0^+$ and $1^+$
states of $(c\bar{s})$ whose masses are thus pushed down.  In short,
the $\mathrm{D}_s(2317)$ and $\mathrm{D}_s(2457)$ states are perhaps
quark--antiquark states mixed with four-quark states and hence lighter
than predicted in simple quark models.  A similar scenario was
suggested, e.g., in
\cite{Browder:2003fk,Terasaki:2005ne,vanBeveren:2003kd}.
\subsection{Mesons with heavy hidden flavour}
There is already a long history of possible multiquarks with hidden
charm, and any new candidate is examined with precaution.  In the late
70's, the molecular charmonium was proposed
\cite{Voloshin:1976ap,DeRujula:1976qd} for high-lying $\psi(4.028)$ or
$\psi(4.414)$ states, to account for their narrowness or for the
deviation from spin counting rules for the relative branching ratios
to $\mathrm{D}\overline{\mathrm{D}}$,
$\mathrm{D}\overline{\mathrm{D}}{}^*+\text{c.c.}$ and
$\mathrm{D}^*\overline{\mathrm{D}}{}^*$.  However, these decay
properties were explained by the node structure of these states being
mostly $(c\bar{c})$ $n{}^3\mathrm{S}_1$ with radial number $n=3$ or
$n=4$ \cite{LeYaouanc:1977ux,LeYaouanc:1977gm,Eichten:1979ms}.

More recently, the states X(3872), X(3940) and Y(4260) were found in
B-factories, and at least the X(3872) was confirmed elsewhere
\cite{Jin:2004ck,Swanson:2006st}.  Scrutinising the mass, decay
properties and production mechanism does not lead to a fully
convincing $n {}^{2s+1}L_J$ assignment in the spectrum of charmonium.
Hence other structures have been proposed.  A
$\mathrm{D}\overline{\mathrm{D}}{}^*+\text{c.c.}$ molecular structure
is particularly tempting for the X(3872), since it lies just above
this threshold, and binding or near binding of this system was
predicted on the basis of long-range nuclear interaction between
$\mathrm{D}$ and $\overline{\mathrm{D}}{}^*$.  For references, see
\cite{Voloshin:2004mh,Tornqvist:2004qy,Swanson:2003tb,Braaten:2005ai,Close:2003sg},
where the X(3872) is analysed, and also the comments by Susuki on this
picture \cite{Suzuki:2005ha}.

Another possibility is that one of the X(3872), X(3940) and Y(4260)
resonances is a state of the long-awaited charmonium hybrid,
schematically denoted $(c\bar{c}g)$.  Already in the 70's and early
80's, one speculated about some new kind of states in mesons and
baryons, where the string linking quarks, or the gluon field in the
QCD language, is excited, see, e.g.,
Ref.~\cite{Giles:1977mp,Horn:1977rq,Hasenfratz:1980jv}, followed by
several studies in the framework of models with constituent gluons or
flux tube, or within QCD sum rules or lattice QCD. The possibility
that one of the new hidden-charm meson is an hybrid is discussed,
e.g., in \cite{Kou:2005gt,Close:2005iz}. The isospin violation observed in the X(3872) is hardly explained in the hybrid scenario. On the other hand, the hybrid interpretation of the Y(4260) is better supported by the data, as stressed in the recent analysis by the CLEO collaboration \cite{Coan:2006rv}.

As neither the molecular nor the hybrid interpretation has won an
overall consensus, yet, the door remains open for a four-quark
interpretation $(c\bar{c}q\bar{q})$ or $(c\bar{c}s\bar{s})$.  Maiani
et al., in particular, suggested a diquark--antidiquark picture of
these states, for instance $(cs)(\bar{c}\bar{s})$ for the Y(4260)
\cite{Maiani:2005pe}.  The effective mass of the diquark, in this
approach, accounts for the strong quark--quark correlation, and the
$(cs)$ subsystem is restricted to have colour $\bar{3}$ and spin
$s=0$.  In the calculation presented here, any subsystem has all the
possible quantum numbers compatible with a given $J^{PC}$ hypothesis
for the whole system, and the weight of each configuration is
determined by diagonalising the chromomagnetic interaction.

As already discussed in \cite{Hogaasen:2005jv}, this procedure
highlights a remarkable eigenstate, which is a pure
$(c\bar{c})_8(q\bar{q})_8$ octet--octet of colour in the limit where
$C_{cq}=C_{c\bar{q}}$, and has a small $(c\bar{c})_1(q\bar{q})_1$
singlet--singlet component if this condition is broken, restricted
however to $(s=1)\otimes (s=1)$ for the spins, explaining the observed
$\jp+\omega$ and $\jp+\pi^+\pi^-$ modes and the absence of
$\jp+\text{pseudoscalar}$.  In the crossed rearrangement
$(c\bar{q})(\bar{c}q)$ of the constituents, the colour content is
dominantly singlet--singlet, but this is
$\mathrm{D}\overline{\mathrm{D}}{}^*+\text{c.c.}$, which is suppressed
by the lack of phase-space. The chromomagnetic mechanism for tetraquarks with hidden charm has been further studied by Stancu~\cite{Stancu:2006st}.

As for the $X_b(b\bar{b} q\bar{q})$, the analogue of $X(3872)$ in the hidden-beauty sector, if the parameters are tuned to fit 
the measured values of the masses of B, B${}^*$, $\Upsilon$ and
$\Lambda_b$ hadrons, it appears to be stable
against dissociation into $\mathrm{B}\overline{\mathrm{B}}{}^*$, with, however, possible decay into $\Upsilon+\omega$~\cite{Hogaasen:2005jv}. 
%
\subsection{Mesons with double heavy flavour}
It often happens in the field of exotic hadrons that a new
calculation of a given configuration gives results at variance or in
serious conflict with the previous ones. An example is pentaquark
with flavour content $(uudd\bar{s})$ or its heavier analogue with $s$
replaced by $c$, found either unbound or nearly stable, and in the
latter case, either with positive or negative parity. An exception is
the sector of $(QQ'\bar{q}\bar{q'})$ states with two heavy quarks and
two light antiquarks,  or \textsl{vice-versa}. An abundant literature
has been accumulated over the years for these exotic configurations
\cite{Ader:1981db,Zouzou:1986qh,Heller:1986bt,Carlson:1988hh,Lipkin:1986dw,%
Brink:1998as,Gelman:2002wf,Janc:2004qn}.  All the papers convey
basically the same message, namely if the mass ratio $M/m$ of quarks
to antiquarks is large enough, this state becomes stable against
spontaneous dissociation into $(Q\bar{q}) + (Q'\bar{q}')$, which is
the lowest threshold if $Q'$ is heavier than $Q$ and $q'$ than $q$.
If $Q\neq Q'$, $M$ can be taken as twice their reduced mass.

This binding occurs for a spin- and flavour-independent
interaction, and can be said to be of chromoelectric nature. The same
effect is observed in atomic physics (see, e.g., \cite{Armour:2005}
for references) for a system of four unit charges
$(M^+,M^+,m^-,m^-)$: whilst the systems with equal masses, which
corresponds to the positronium molecule Ps$_2$, is just by a small
amount below the threshold for dissociation into two positronium
atoms, the hydrogen molecule, for which $M\gg m$, is more deeply
bound and possesses a rich spectrum of excitations.

In model calculations, mesons with double charm, corresponding to the
case of $Q=Q'=c$, appear at best as marginally bound
\cite{Janc:2004qn}, and are most often found to be unbound.  It is
difficult to say whether the ground-state is actually unstable in the
assumed model, a better treatment of the 4-body problem might change
the conclusions.  It is thus important to analyse to which extent the
chromomagnetic interaction can help achieving the binding, or has a
repulsive effect.

For $(QQ\bar{q}\bar{q})$ with identical heavy quarks, the
chromomagnetic interaction is optimal for $J^P=1^+$, since the Pauli
principle forbids the $0^+$ eigenstates with the lowest eigenvalue of
$H_{\text{CM}}$.  This restriction does not apply to states with charm
and beauty, and the most favourable situation occurs to be
$(bc\bar{u}\bar{d})$ with $J^P=0^+$ and isospin $I=0$.  The very large
value of the mass ratio $M_Q/m_n$, where $M_Q^{-1}$ is the average of
the inverse masses $m_c$ and $m_b$\footnote{The inverse masses,
entering linearly the Schr{\"o}dinger equation are more pertinent that
the masses themselves to follow the variation of the binding energy},
presumably gives binding or almost binding from the sole
chromoelectric effects.  The chromomagnetic interaction is also
favourable, and, if alone, would give a binding of more than 100 MeV.

Of course, it would be naive to add this chromomagnetic binding
energy to the chromoelectric binding estimated in another model. Each
term, to provide with the best  attraction, require a specific
internal spin--colour function, and with the two interaction terms
switched on, a  compromise has to be found. It is however
reasonable to believe that the net energy would lie below the lowest
of the chromoelectric and the chromomagnetic ones.

A search for this exotic meson with charm $C=1$ and beauty $B=-1$ does
not seem out of reach.  The ground state of $(b\bar{c})$ has already
been found \cite{Eidelman:2004wy} and the ground state of $(ccq)$ has
been observed in an experiment
\cite{Mattson:2002vu,Ocherashvili:2004hi}, though not confirmed in
others \cite{Aubert:2006qw}.  Moreover, double-charm production has
been observed at beauty factories, permitting in particular to
identify new charmonia recoiling against the $\jp$ \cite{Choi:2002na}.
Hence other systems with two units of heavy flavour should be
accessible.  Also, the potential of heavy-ion collisions as a flavour
factory to produce exotic hadrons \cite{Kabana:2004hh} has not yet
been fully exploited.
\section{Outlook}\label{se:con}
A survey of various flavour configurations show that the spectrum of
the lowest positive-parity mesons is greatly influenced by multiquark
configurations.  For the crypto-exotic states, a detailed comparison
with experimental data cannot avoid a delicate mixing scheme involving
orbitally excited quark--antiquark states (including radial
excitation) and hybrids, and presumably glueball for the case of zero
isospin and strangeness.  The analysis presented here shows that the
multiquark component has a level ordering somewhat different of the
one currently used, with scalar $(q\bar{q}s\bar{s})$ mesons higher in
the spectrum.  The possibility of a spin $S=2$, isospin $I=2$
resonance should be considered seriously.  There are, indeed,
indications\cite{Anikin:2005ur} that this exotic state might be in the
recent experimental data of the L3 collaboration \cite{Achard:2005db}, and
in earlier $\gamma\gamma$ data in gamma-gamma reactions
\cite{Li:1981ez,Liu:1987dd,Achasov:1981kh,Achasov:1987ak}

In the sector of naked charm $C=1$, the dynamical scheme presented
here does not permit to assign the $\mathrm{D}_{s,J}$ states to be the
$J^P=0^+$ and $1^+$ states of $(c\bar{s}n\bar{n})$.  A tricky
possibility, however, exists, that these multiquarks are broad but lie
not too far and hence that the $0^+$ and $1^+$ states of $(c\bar{s})$
are pushed down by mixing.

In the sector of charm $C=2$, or in the sector with charm $C=1$ and
beauty $B=-1$, a serious multiquark candidate is $(bc\bar{u}\bar{d})$
with $J^P=1^+$. The main reason is the tendency for heavy quarks to
experience the best binding in a given --flavour independent --
confining well, i.e., the binding is essentially chromoelectric.
However, the chromomagnetic interaction helps.

In the more accessible sector with hidden charm, the X(3872) is well
described as an eigenstate of the chromomagnetic interaction, which is
basically a pure octet--octet $[(c\bar{c})_8(q\bar{q})_8]$, and a
small impurity which explains the data on the $\jp\pi^+\pi^-$ decay.
In the other coupling scheme, the X(3872) is largely of
singlet--singlet type, i.e., $[(c\bar{q})_1(\bar{c}q)_1$, but as
$\mathrm{D}\overline{\mathrm{D}}$ is forbidden, the decay proceeds via
$\mathrm{D}\overline{\mathrm{D}}{}^*+\text{c.c.}$ and lacks
phase-space.  This explains the remarkably small width of this state.
This picture raises interesting questions.  The quark model predicts
degenerate isospin states, $I=0$ and $I=1$.  The neutral members mix
by annihilation diagrams, and the narrowest shows up most strikingly
in the data.  The charged members of the $I=1$ multiplet are perhaps
not so easily produced as the neutral in $\mathrm{B}$ decay, and
should be searched for by other production mechanisms.  The direct
treatment of the four-quark interaction differs from the
nuclear-physics approach, which used the well identified $\mathrm{D}$
and $\mathrm{D}{}^*$ mesons, and the well known Yukawa mechanism of
long-range interaction between hadrons.  However, in absence of
evidence for a repulsive core keeping the hadrons well separated, the
situation is different from that of nucleons within a nucleus, and it
is not sure whether or not the bound states dynamics remains dominated by
long-range forces.

It is worth stressing once more that a four-quark explanation of a few
remarkable mesons such as the $X(3872)$ does not imply an abundance
of multiquark states in the forthcoming experimental data.  Most
S-wave multiquarks are very broad, as they spontaneously split into
two colour-singlet hadrons.  Only the states with very low mass or
peculiar internal structure survive this dissociation and show up as
narrow peaks.

\begin{acknowledgments}
H.H. thanks the LAPTH, Annecy-le-Vieux, and the CERN Particle Theory
Unit for their kind hospitality.  The comments by M.~Asghar on the
manuscript are gratefully acknowledged.
\end{acknowledgments}

%
%

%
\end{document}